\documentclass[%
 reprint,
 amsmath,amssymb,
 aps,
 prx,
longbibliography
]{revtex4-2}

\usepackage{graphicx}
\usepackage[dvipsnames]{xcolor}
\usepackage{dcolumn}
\usepackage{bm}
\PassOptionsToPackage{hyphens}{url}
\usepackage{hyperref} 
\hypersetup{colorlinks=true, linkcolor=blue, citecolor=ForestGreen, urlcolor=DarkOrchid}
\usepackage{dsfont}

\usepackage{defs}

\begin{document}

\title{Quantum combinatorial optimization beyond the variational paradigm: simple schedules for hard problems}

\author{Tim Bode$^{1}$, Krish Ramesh$^{1}$, and Tobias Stollenwerk$^{1}$}
\affiliation{%
 $^{1}$Institute for Quantum Computing Analytics (PGI-12), Forschungszentrum Jülich, 52425 Jülich, Germany
}%

\date{\today}

\begin{abstract}
Advances in quantum algorithms suggest a tentative scaling advantage on certain combinatorial optimization problems. Recent work, however, has also reinforced the idea that barren plateaus render variational algorithms ineffective on large Hilbert spaces. Hence, finding annealing protocols by variation ultimately appears to be difficult. Similarly, the adiabatic theorem fails on hard problem instances with first-order quantum phase transitions. Here, we show how to use the spin coherent-state path integral to shape the geometry of quantum adiabatic evolution, leading to annealing protocols at polynomial overhead that provide orders-of-magnitude improvements in the probability to measure optimal solutions, relative to linear protocols. These improvements are not obtained on a controllable toy problem but on randomly generated hard instances (Sherrington-Kirkpatrick and Maximum 2-Satisfiability), making them generic and robust. Our method works for large systems and may thus be used to improve the performance of state-of-the-art quantum devices.
\end{abstract}

\maketitle


\section{\label{sec:intro}Introduction}

One of the current challenges for quantum combinatorial optimization presents itself like this: on one hand, even if there is a mild polynomial advantage of quantum algorithms over the best classical ones~\cite{ebadiQuantumOptimizationMaximum2022, zancaQuantumAnnealingSpeedup2016}, higher operating costs might still render this practically unviable. On the other hand, though it may take a fault-tolerant quantum computer to achieve, the prospect of having a ``universal sampler'', capable of discovering unlikely low-energy states, is rather fascinating. Granted, we know that in worst-case scenarios, this will be \textit{exponentially} unlikely~\cite{aminFirstorderQuantumPhase2009, knyshZerotemperatureQuantumAnnealing2016}. Still, it is simply not clear if there is an intermediate hardness regime where a quantum device could be capable of sometimes finding solutions that no other method can. This leads to a simple question: Are there ways to systematically increase the likelihood of optimal solutions? In this work, we present one such method.

Progress in the use of large-scale quantum annealers~\cite{kingQuantumCriticalDynamics2023} suggests that it might be possible to approach quantum advantage on certain types of combinatorial optimization problems~\cite{au-yeungNPhardNoLonger2023}. Similar perspectives have been given in the context of the Quantum Approximate Optimization Algorithm (QAOA)~\cite{shaydulinEvidenceScalingAdvantage2024, boulebnaneSolvingBooleanSatisfiability2024, marwahaBoundsApproximatingMax2022}; multistage quantum walks, in turn, have been shown to be able to outperform the QAOA in certain cases~\cite{gerblichAdvantagesMultistageQuantum2024}. All of these results, however, are put in perspective by recent investigations showing that noise effectively truncates quantum circuits to logarithmic depth~\cite{meleNoiseinducedShallowCircuits2024}; it also rules out exponential speed-ups in expectation-value estimation~\cite{quekExponentiallyTighterBounds2024}. Recent work has also reinforced the idea that the barren-plateau phenomenon will render variational quantum algorithms ineffective on large Hilbert spaces~\cite{kaziAnalyzingQuantumApproximate2024, cerezoDoesProvableAbsence2024}. Hence, finding useful variational annealing protocols (digital or analog) for quantum combinatorial optimization appears to be difficult. Naturally, one could in theory always reach the ground state thanks to the existence of the adiabatic theorem, were it not for the fact that, once again, hard problem instances, which are characterized by first-order quantum phase transitions~\cite{aminFirstorderQuantumPhase2009}, will require impractically long annealing times. 
While diabatic strategies, such as excitation followed by return to the ground state~\cite{crossonDifferentStrategiesOptimization2014} or adiabatic gauge potentials~\cite{kolodrubetzGeometryNonadiabaticResponse2017}, are known to sometimes help at least for small systems or short evolution times, respectively, it seems improbable that such effects will persist for large Hilbert spaces with complex energy landscapes. If we accept the outlook that optimal schedules for large, hard instances cannot be reached in practice, again the question becomes if there are other methods~\cite{bishopSetAnnealingProtocols2023, finzgarDesigningQuantumAnnealing2024} to obtain schedules that increase the likelihood of obtaining low-energy states.

{ Practical applications of discrete optimization are manifold and have been widely discussed in the literature~\cite{yarkoniQuantumAnnealingIndustry2022a, abbasChallengesOpportunitiesQuantum2024}. Prominent examples are use cases in financial~\cite{orusQuantumComputingFinance2019} and energy markets~\cite{braunOptimizationUncertaintyFundamental2023, halffmannQuantumComputingApproach2023}. Hard problems in biology such as protein folding~\cite{perdomo-ortizFindingLowenergyConformations2012} have also been addressed.}

Our main contribution to the questions we have raised is the following. By combining the geometric approach to adiabatic quantum computation (AQC)~\cite{rezakhaniIntrinsicGeometryQuantum2010} with the spin coherent-state path integral for quantum spin glasses~\cite{stoneSemiclassicalPropagatorSpin2000, misra-spieldennerMeanFieldApproximateOptimization2023, bodeAdiabaticBottlenecksQuantum2024}, we show in detail how semi-classical observables can be used to systematically derive ``adaptive'' annealing schedules at $\mathrm{poly}(N)$ overhead, where $N$ is the problem size. Such schedules can lead to orders-of-magnitude improvements in the AQC probability $P$ to observe the optimal solution of a given problem instance. We work with two established hard-instance data sets { with known \textit{exact} ground states}: one for the Sherrington-Kirkpatrick (SK) model as used in Ref.~\cite{bodeAdiabaticBottlenecksQuantum2024}, the other for Maximum 2-Satisfiability (MAX 2-SAT) as introduced earlier by Shor \textit{et al.}~\cite{crossonDifferentStrategiesOptimization2014}. On the latter data, we improve the geometric mean success probability across all instances from $ 0.01\%$ in the linear-schedule case to $3.1\%$ with our method; for single instances, the improvement can reach more than four orders of magnitude (cf.~section~\ref{subsec:MAX2SAT_results}). For the SK model, we find that our schedules are close to the ``ideal'' ones derived from the exact adiabatic bottleneck.

We emphasize two points: First, the improvements we see are not obtained on a specific controllable toy problem, but on randomly generated hard instances of two different problems, which leads us to expect that these improvements are more generic and robust than could be deduced from the former. Second, our method works for large system sizes without modification, which means it is ready to be used to improve the results of state-of-the-art quantum devices with larger numbers of qubits.

This article is organized as follows. After defining our framework for quantum optimization in section~\ref{sec:qcopt} alongside our semi-classical theory, we show how to combine it with the geometric approach to AQC in section~\ref{sec:geom_sched}. The two data sets of hard instances to which these methods will be applied are then briefly described in section~\ref{sec:instances}, after which we present our results in section~\ref{sec:results}, followed by concluding remarks in section~\ref{sec:conc}. All problem instances and the corresponding results are publicly available~\cite{bodeDataQuantumCombinatorial2024}.

\section{Quantum Combinatorial Optimization}\label{sec:qcopt}

For the purposes of this work, we do not differentiate between AQC and the QAOA in the sense that one may consider { the Trotterized version of continuous-schedule AQC a limiting case of} fixed-angle QAOA. Therefore, after defining the continuous-time adiabatic Hamiltonian below, we show once how to initialize the corresponding QAOA angles based on a given annealing protocol. In our theoretical derivations, we then stick to a formulation in terms of continuous time, while we effectively use fixed-angle QAOA in our numerics.

\subsection{\label{subsec:prob_def} Problem Definition}

\begin{figure}[htb!]
\begin{center}
\includegraphics[width=\linewidth]{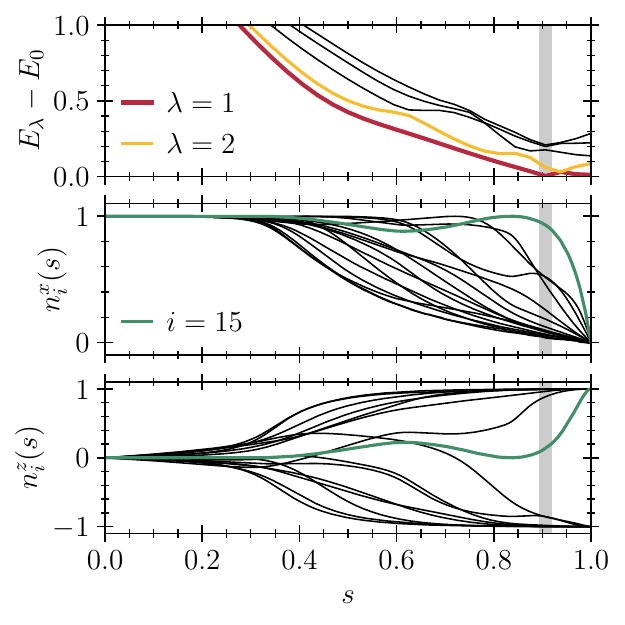}\vspace{-0.5cm}
\caption{\label{fig:mean_fields_N_17_seed_9129} Exemplary SK instance at $N=17$. The top panel shows the exact spectrum. The lower panels display the mean-field trajectories, with the most frustrated spin highlighted. The vertical bar indicates the position of the adiabatic bottleneck $s_*$.}
\end{center}
\end{figure}

It is well-known that (quadratic) combinatorial optimization problems can be represented via an Ising Hamiltonian $\hat H_Z$, to which we add a transverse-field Hamiltonian $\hat H_X$, where
\begin{align}\label{eq:H_Pand_H_D}
\hat H_Z = - \sum_{i=1}^N \bigg[ h_i  + \sum_{j>i} J_{ij}  \hat Z_j \bigg] \hat Z_i, \quad \hat H_X = - h_X\sum_{i=1}^N \hat X_i,
\end{align}
and we set $\hbar=1$. The interaction matrix $J_{ij}$ and the local magnetic fields $h_i$ then define the given problem instance in such a way that the ground state of $\hat H_Z$ can be identified with the optimal solution of the corresponding optimization problem. To fix time and frequency units, we set $h_X = 1$ everywhere. The evolution time is defined by letting $ 0 \leq t \leq T$. The total Hamiltonian is
\begin{align}\label{eq:total_ham}
    \hat H(\boldsymbol{s}(t)) = s^1(t)\hat H_X + s^2(t) \hat H_Z,
\end{align}
where $\boldsymbol{s}(t) =  (s^1(t), s^2(t))$ defines the (two-component) schedule parameterizing the change from the initial Hamiltonian $\hat H_X$ to the final Hamiltonian $\hat H_Z$, i.e.\ we assume boundary conditions $\boldsymbol{s}(0) =  (1, 0)$ and $\boldsymbol{s}(T) =  (0, 1)$. The simplest possible choice is the (one-component) linear schedule given by 
\begin{align}
    \begin{split}
        s^1(t) = 1 - t/T,  \quad s^2(t) = t/T.
    \end{split}
\end{align}
The general schedule $\boldsymbol{s}(t)$ can be converted to QAOA angles in a straightforward way. Defining the usual $p$-layer QAOA state as
\begin{align}
    |\boldsymbol{\beta}, \boldsymbol{\gamma}\rangle = \hat U_{X}\left(\beta_{p}\right) \hat U_Z\left(\gamma_{p}\right) \cdots \hat U_{X}\left(\beta_{1}\right) \hat U_Z\left(\gamma_{1}\right) |+\rangle^{\otimes N}, 
\end{align}
with the unitaries $\hat U_{X}(\beta_k) = \ee^{-\ii \beta_k \hat H_X}$ and $\hat U_{Z}(\gamma_k) = \ee^{-\ii \gamma_k \hat H_Z}$ for $k = 1, ..., p$, if one employs a simple lowest-order Trotterization, the angles become
\begin{align}\label{eq:qaoa_angles}
    \beta_k = \tau s^1(Tk/p), \quad \gamma_k = \tau s^2(Tk/p), \quad \tau = T/p.
\end{align}
It is, of course, possible to instead use higher-order Trotterization schemes~\cite{willschBenchmarkingQuantumApproximate2020} to derive $\boldsymbol{\beta}, \boldsymbol{\gamma}$ from $\boldsymbol{s}(t)$, which we do below whenever applying linear schedules. An example of a typical energy spectrum featuring the finite-size analog of a first-order phase transition is shown in Fig.~\ref{fig:mean_fields_N_17_seed_9129} for the SK model.

\subsection{Classical Equations of Motion}\label{subsec:mf_eom}

The derivation of the mean-field approximation has been detailed in Ref.~\cite{misra-spieldennerMeanFieldApproximateOptimization2023}. The mean-field form of the adiabatic Hamiltonian~\eqref{eq:total_ham} is given by
\begin{align}\label{eq:mf_H}
    \begin{split}
        H(t) = &-\sum_{i=1}^N \Bigg[ s^1(t) n^x_i + s^2(t) \bigg(h_i + \sum_{j>i} J_{ij}  n^z_j\bigg) n^z_i\Bigg].
    \end{split}
\end{align}
The classical equations of motion in our present notation then read
\begin{align}\begin{split}\label{eq:mf_eom}
\dot n^x_i(t) &= \phantom{-}2 s^2(t) m_i(t) n^y_i(t),\\
\dot n^y_i(t) &= -2 s^2(t) m_i(t) n^x_i(t) + 2 s^1(t)  n^z_i(t) , \\
\dot n^z_i(t) &= -2 s^1(t)  n^y_i(t),
\end{split}\end{align}
where we have introduced the local magnetic field
\begin{align}\begin{split}\label{eq:magnetization}
m_i(t) = h_i + \sum_{j=1}^N J_{ij} n^z_j(t).
\end{split}\end{align}
Recently, it was brought to our attention that a similar model had been investigated ten years ago~\cite{smolinClassicalSignatureQuantum2014, shinHowQuantumDWave2014} in the context of a discussion around the ``quantumness'' of earlier quantum annealers. These earlier models do not include the $y$-components of the classical spins explicitly, and are only equivalent to ours in the adiabatic limit (i.e.\ for $T\to\infty$), while they generally differ for finite $T$. The main difference, however, is that our more general model allows for the derivation of the semi-classical fluctuations~\cite{misra-spieldennerMeanFieldApproximateOptimization2023} around the mean-field trajectories, which we present in the next section.

\subsection{Gaussian Fluctuations}
\label{subsec:flucs}

The fluctuations of trajectories on the Bloch sphere can be shown to be described by a flat path-integral measure to one-loop order~\cite{stoneSemiclassicalPropagatorSpin2000}. Instead of this well-defined Gaussian spin coherent-state path integral~\cite{misra-spieldennerMeanFieldApproximateOptimization2023}, one can also employ the quadratic Hamiltonian
\begin{align}\label{eq:para_ham}
    \begin{split}
        \hat{\mathcal{H}}(t) = \frac 12 \sum_{i,j=1}^N\Big[ &A_{ij}(t)\hat\eta_i^\dagger \hat\eta_j^{\phantom{\dagger}} + A_{ji}(t)\hat\eta_j^\dagger \hat\eta_i^{\phantom{\dagger}} \\[-2mm]
        + &B_{ij}(t)\hat\eta_i^\dagger \hat\eta_j^\dagger + \bar{B}_{ji}(t) \hat\eta_i^{\phantom{\dagger}}\hat\eta_j^{\phantom{\dagger}} \Big]
    \end{split}
\end{align}
to define the corresponding quasi-particles~\cite{bodeAdiabaticBottlenecksQuantum2024} with creation and annihilation operators $\hat\eta_i^{{\dagger}}$, $\hat\eta_i^{\phantom{\dagger}}$. Since Hamiltonians in the context of combinatorial optimization typically have frustrated ground states, { their quasi-particles are usually called \textit{paramagnons} (as opposed to, e.g., the quasi-particle excitations on top of a ferromagnetically ordered ground state, which would be addressed simply as \textit{magnons})}. 

The diagonal coefficients of Eq.~\eqref{eq:para_ham} are now given by $B_{ii}(t) = 0$ and
\begin{align}
\begin{split}
\label{eq:AB_diag}
A_{ii}(t) =  \frac{2 s^1(t) n^x_i(t)}{1 + {\sigma}^*_i n^z_i(t)} + 2 s^2(t) {\sigma}^*_i m_i(t),
\end{split}
\end{align}
where the vector $\boldsymbol{\sigma}^* = \left(\mathrm{sign}(n_1^z(T), ..., \mathrm{sign}(n_N^z(T) \right)^T$ is the mean-field solution of the combinatorial problem instance. It is obtained by evolving the classical equations~\eqref{eq:mf_eom} up to time $T$. For the off-diagonal coefficients, one finds
\begin{align}\begin{split}
\label{eq:AB_off-diag}
A_{ij}(t) &= -s^2(t)\, J_{ij} n^+_i(t) n^-_j(t),\\
B_{ij}(t) &= \phantom{-}s^2(t)\, J_{ij} n^+_i(t) n^+_j(t),
\end{split}\end{align}
with $n_i^{\pm}(t)= {\sigma}^*_i n^x_i(t) \pm i n^y_i(t)$. 

\begin{figure}[htb!]
\begin{center}
\includegraphics[width=\linewidth]{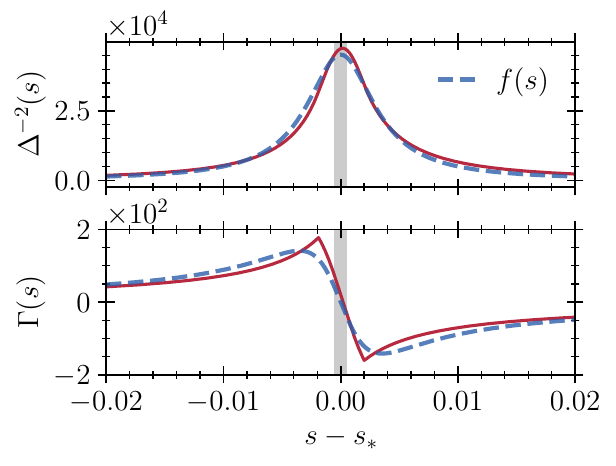}\vspace{-0.5cm}
\caption{\label{fig:exact_gap_N_17_seed_9129} The effective adiabatic metric Eq.~\eqref{eq:approx_metric} together with a scaled Cauchy distribution~\eqref{eq:lorentzian} for the example of Fig.~\ref{fig:mean_fields_N_17_seed_9129}. The corresponding Christoffel symbols~\eqref{eq:damping} are shown in the lower panel. Note that replacing the Cauchy distribution by a Gaussian (which has ``thinner'' tails) is another option that we found to yield similarly good results.}
\end{center}
\end{figure}

To be able to model the non-equilibrium dynamics of the paramagnons, we introduce the {greater} and {lesser} Green functions~\cite{Rammer_2007} defined by
\begin{subequations}\label{eq:greater_lesser_ops}
    \begin{align}
        G_{ij}^>(t, t') &= -\ii\langle \hat\eta_i^{\phantom{\dagger}}(t) \hat\eta_j^{{\dagger}}(t') \rangle, \\
        G_{ij}^<(t, t') &= -\ii\langle \hat\eta_j^{{\dagger}}(t') \hat\eta_i^{\phantom{\dagger}}(t) \rangle.
    \end{align}
\end{subequations}
The semi-classical equations of motion of these propagators are 
\begin{subequations}\label{eq:G_g_l}
\begin{align}
\label{eq:G_t}
        \left[i\sigma_3\overrightarrow{\partial_t} - \mathcal{H}(t)\right]\boldsymbol{G}^\gtrless(t, t') &= 0, \\
\label{eq:G_tp}
        \boldsymbol{G}^\gtrless(t, t')\left[-i\sigma_3\overleftarrow{\partial_{t'}} - \mathcal{H}(t')\right] &= 0,
\end{align}
\end{subequations}
where
\begin{align}
\label{eq:H_t}
\mathcal{H}(t) = \begin{pmatrix}
 A(t) &  B(t) \\ 
 B^\dagger(t) & \bar{A}(t)
\end{pmatrix}, \quad
\sigma_3 =
\begin{pmatrix}
\mathds{1} & \mathds{0}  \\  & -\mathds{1}
\end{pmatrix}.
\end{align}
In this work, we only need the statistical information about the fluctuations at equal times, for which reason it is sufficient to consider the so-called \textit{statistical} function
\begin{align}\label{eq:stat_func}
    \begin{split}
        \boldsymbol{F}(t, t') &= \sbs{\boldsymbol{G}^>(t, t') + \boldsymbol{G}^<(t, t')} \sigma_3,
    \end{split}
\end{align}
which at equal times $t'=t$ has the convenient property
\begin{align}\label{eq:paramagnon_number}
    \ii F_{ii}(t, t) = 2\mathcal{N}_i(t) + 1,
\end{align}
where $\mathcal{N}_i(t)$ can be interpreted as a time-dependent particle number. Its equation of motion at equal times follows from Eqs.~\eqref{eq:G_g_l} as
\begin{align}\label{eq:eom_stat_GF}
    \ii\partial_t \boldsymbol{F}(t, t) &= \left[\sigma_3 \mathcal{H}(t), \boldsymbol{F}(t, t)\right].
\end{align}

Together with the initial condition $\boldsymbol{F}(0, 0) = -\ii \sigma_3$, Eq.~\eqref{eq:eom_stat_GF} defines a $2N\times 2N$ initial-value problem. Regarding the algorithmic overhead of our method, this $\sim N^2$ cost is the biggest contribution since the number of mean-field equations of motion~\eqref{eq:mf_eom} scales as $\sim N$. Once Eq.~\eqref{eq:eom_stat_GF} is solved, the additional overhead of obtaining the corresponding annealing schedules, which we describe in detail in section~\ref{sec:geom_sched}, is actually \textit{independent} of the system size $N$. In total, then, we have an algorithmic overhead that scales as $\mathrm{poly}(N)$.

\section{\label{sec:geom_sched} Geometric Approach to Optimal Schedules}

For hard instances, the slow dynamics of the total Hamiltonian~\eqref{eq:total_ham} will be dominated by at least one exponentially small bottleneck in the instantaneous spectral gap~\cite{knyshZerotemperatureQuantumAnnealing2016} between the ground and first excited states, which we write as $\Delta(\boldsymbol{s}(t)) = E_1(\boldsymbol{s}(t)) - E_0(\boldsymbol{s}(t))$. Assuming for simplicity a non-degenerate spectrum, we also introduce the adiabatic \textit{metric tensor}~\cite{rezakhaniIntrinsicGeometryQuantum2010}
\begin{align}\label{eq:gap_metric}
    \begin{split}
        g_{\mu\nu}(\boldsymbol{s}) = \mathrm{Re}\hspace{-0.0mm}\sum_{\lambda>0}\hspace{-0.0mm} \frac{\langle \lambda|\partial_\mu \hat H(\boldsymbol{s})  |0\rangle \hspace{-0.75mm}\langle 0|\partial_\nu \hat H(\boldsymbol{s})  |\lambda\rangle}{(E_\lambda(\boldsymbol{s}) - E_0(\boldsymbol{s}))^2},
    \end{split}
\end{align}
where $\partial_{\mu} := \partial/\partial s^{\mu}$, and we have left the parameter dependencies of the instantaneous eigenstates implicit. Note also that Eq.~\eqref{eq:gap_metric} will typically be dominated by the first excited state $m=1$, i.e.\ by the inverse of the minimal gap $\Delta(\boldsymbol{s}(t))$.

The geometric approach to adiabatic quantum computation outlined in Ref.~\cite{rezakhaniIntrinsicGeometryQuantum2010} now allows one to derive optimal schedules from a geodesic equation. From the adiabatic metric tensor given in Eq.~\eqref{eq:gap_metric}, one defines the line integral
\begin{align}
    \begin{split}
        \mathcal{L}(\boldsymbol{s}(t)) \sim \int_0^t \mathrm{d}t' \sqrt{g_{\mu\nu}(\boldsymbol{s})\dot{s}^\mu(t')\dot{s}^\nu(t')},
    \end{split}
\end{align}
which leads to the usual geodesic equation
\begin{align}
    \begin{split}
        \ddot{s}^\mu(t) &= -\Gamma^\mu_{\nu\xi}\dot{s}^\nu(t)\dot{s}^\xi(t),
    \end{split}
\end{align}
with Christoffel symbols defined as
\begin{align}
    \begin{split}
        \Gamma^\mu_{\nu\xi} &= \frac{1}{2}g^{\mu\eta}\left( \partial_\xi g_{\eta \nu} + \partial_\nu g_{\eta \xi} - \partial_\eta g_{\nu \xi}\right).
    \end{split}
\end{align}
In the one-parameter case to which we will specialize below, we then have 
\begin{align}
    \begin{split}
        s^1(t) = 1 - s(t),  \qquad s^2(t) = s(t),
    \end{split}
\end{align}
whence the geodesic equation simplifies to
\begin{align}\label{eq:geodesic}
    \begin{split}
        \ddot{s}(t) &= -\Gamma(s)\dot{s}^2(t)
    \end{split}
\end{align}
with boundary conditions $s(0) = 0$, $s(1) = 1$. The Christoffel symbol in Eq.~\eqref{eq:geodesic} is
\begin{align}\label{eq:damping}
    \Gamma(s) = g^{-1}(s) g'(s)/2,
\end{align}
where { $g(s)$ is a one-parameter metric ``tensor'' and} the prime denotes differentiation. Note that due to these, Eq.~\eqref{eq:geodesic} is not an initial-value problem but rather a second-order \textit{boundary-value} problem (BVP), which can be solved numerically, e.g., by means of the monotonic implicit Runge-Kutta (MIRK) or the shooting method~\cite{rackauckas2017differentialequations, bezanson2017julia}. 

For hard instances, { the metric} will almost always be dominated strongly by the energy gap, i.e.~we can assume
\begin{align}\label{eq:approx_metric}
    \begin{split}
        g(s) &\simeq \Delta^{-2}(s),
    \end{split}
\end{align}
which then also defines the adiabatic bottleneck via
\begin{align}\label{eq:bottleneck}
    s_* := \underset{s\in [0, 1]}{\arg\,\max}\;g(s).
\end{align}
Inserting Eq.~\eqref{eq:approx_metric} into~\eqref{eq:damping}, one obviously finds 
\begin{align}\label{eq:gamma_ex}
    \Gamma(s) \simeq - \Delta^{-1}(s)\Delta'(s),
\end{align}
an example of which is shown in Fig.~\ref{fig:exact_gap_N_17_seed_9129} together with a scaled Cauchy distribution (Lorentzian)
\begin{align}\label{eq:lorentzian}
    f(s) \sim \frac{\sigma}{(s - s_*)^2 + \sigma^2}
\end{align}
that is a natural approximation to the exact data. 

\begin{figure}[htb!]
\begin{center}
\includegraphics[width=\linewidth]{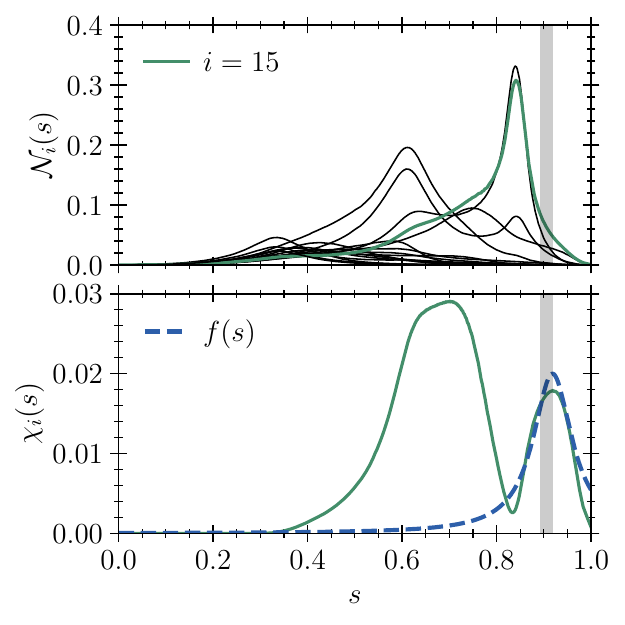}\vspace{-0.5cm}
\caption{\label{fig:fluctuations_N_17_seed_9129} Semi-classical observables for the same $N=17$ example instance as in Fig.~\ref{fig:mean_fields_N_17_seed_9129}. Notice how the paramagnon number $\mathcal{N}_{i}(s)$ of the highlighted spin spikes when approaching the bottleneck (upper panel). In the lower panel, while the localization susceptibility indeed shows two peaks, it is still straightforward to infer the bottleneck position by also taking into account the information from $\mathcal{N}_{i}(s)$. The dashed line finally indicates the approximate Lorentzian~\eqref{eq:lorentzian} that goes into Eq.~\eqref{eq:geodesic} to derive the corresponding adaptive schedule.}
\end{center}
\end{figure}

Note also that a good intuition for Eq.~\eqref{eq:geodesic} can be obtained by thinking of the Christoffel symbol $\Gamma(s)$ as a ``damping'' coefficient: the ``deceleration'' is largest before, and goes to zero at, the adiabatic bottleneck; after passing through, the ``damping'' coefficient becomes \textit{negative}, which results in ``acceleration''. In this mechanical analogy, the ``velocity'' corresponds to the degree of diabaticity with which the Hamiltonian~\eqref{eq:total_ham} is altered, and it is expected to be lowest around the bottleneck.

\subsection{\label{subsec:adap_sched} Adaptive Schedules}

Earlier work~\cite{bodeAdiabaticBottlenecksQuantum2024} indicates that it is possible to obtain an educated guess for the adiabatic bottleneck $s_*$ from the Gaussian fluctuations around the classical mean-field trajectories, an example of which is shown in Fig.~\ref{fig:mean_fields_N_17_seed_9129}. From the $z$-components of these trajectories, we define the local Edwards-Anderson order parameter as 
\begin{align}\label{eq:local_EA_param}
    \begin{split}
        q_i(s) := n^z_i(s)^2,
    \end{split}
\end{align}
which can be thought of as indicating the degree of localization of a given classical trajectory. An example of the semi-classical paramagnon numbers $\mathcal{N}_{i}(s)$ is shown in the upper panel of Fig.~\ref{fig:fluctuations_N_17_seed_9129}. As was already shown in Ref.~\cite{bodeAdiabaticBottlenecksQuantum2024}, the { paramagnon number of spins with high frustration (cf.~Fig.~\ref{fig:mean_fields_N_17_seed_9129}) tends} to grow around the bottleneck $s_*$, which one can understand physically by associating it with the finite-size analog of a first-order phase transition~\cite{aminFirstorderQuantumPhase2009}. If a frustrated spin tries to localize and starts to fluctuate strongly while doing so, this turns out to be a very good semi-classical indicator of the adiabatic bottleneck, which we quantify by combining the local Edwards-Anderson order parameter with the paramagnon number to form the so-called \textit{localization susceptibility}
\begin{align}\label{eq:loc_suscep}
    \begin{split}
        \chi_i(s) = q_i(s) \rbs{1 + \left|z_i(s)\right|^2}^2 \mathcal{N}_{i}(s),
    \end{split}
\end{align}
where 
\begin{align}
\label{eq:z_via_n}
z_i(s) = \frac{n^x_i(s) \pm \ii n^y_i(s)}{1 \pm n^z_i(s)}
\end{align}
follows from the stereographic projection of the Bloch sphere to the complex plane~\cite{misra-spieldennerMeanFieldApproximateOptimization2023, stoneSemiclassicalPropagatorSpin2000}. We remark that the behavior seen in the lower panel of Fig.~\ref{fig:fluctuations_N_17_seed_9129} is indeed typical~\cite{bodeAdiabaticBottlenecksQuantum2024}, thus providing us with a method to replace the exact bottleneck~\eqref{eq:bottleneck} with an approximation $\tilde{s}_*$ derived from the maxima of $\chi_i(s)$. Note that multiple peaks as visible in Fig.~\ref{fig:fluctuations_N_17_seed_9129} do not pose fundamental a problem to this strategy, as one may simply perform several quantum circuits for each bottleneck ``candidate''; in most cases, also taking $\mathcal{N}_i(s)$ into direct consideration will render this unnecessary, { in the sense that the degree of frustration of a particular spin trajectory contains information about its relevance for the bottleneck.}

\begin{figure}[htb!]
\begin{center}
\includegraphics[width=\linewidth]{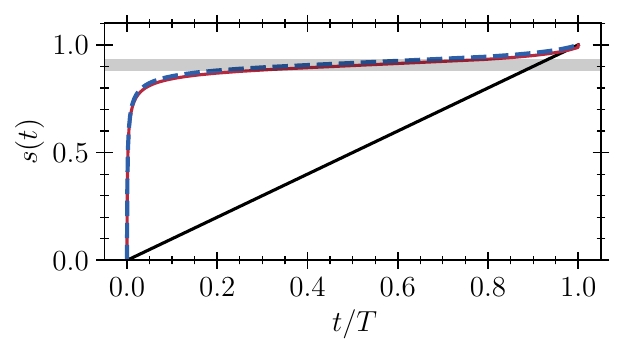}\vspace{-0.5cm}
\caption{\label{fig:schedules_N_17_seed_9129} Adaptive schedule (dashed blue line) for the same SK instance as shown in Fig.~\ref{fig:mean_fields_N_17_seed_9129}. This schedule is derived by solving the boundary-value problem in Eq.~\eqref{eq:geodesic} with the ``force'' term~\eqref{eq:approx_gamma} centered on the approximate bottleneck position $\tilde s_*$ corresponding to the second peak in the lower panel of Fig.~\ref{fig:fluctuations_N_17_seed_9129}. The ``ideal'' schedule following from the same procedure with $\tilde s_*$ replaced by the exact $s_*$ is indicated by the solid (red) line. The parameters of Eq.~\eqref{eq:approx_gamma} used for both are $\tilde\Gamma_0 = 3.20$ and $\sigma = 0.05$. The linear schedule is also drawn for comparison.}
\end{center}
\end{figure}

The Cauchy distribution for the bottleneck in Eq.~\eqref{eq:lorentzian} then leads to a ``force''
\begin{align}\label{eq:approx_gamma}
    \tilde\Gamma(s) \sim -\tilde\Gamma_0 \frac{s - \tilde{s}_*}{(s - \tilde{s}_*)^2 + \sigma^2},
\end{align}
where we have introduced an additional parameter $\tilde\Gamma_0$ by hand. Both parameters $\tilde\Gamma_0$ and $\sigma$ can be tuned to adjust the resulting annealing schedules to external specifications, with $\tilde\Gamma_0$ controlling the strength of the ``force'' from the perspective of our mechanical analogy, while $\sigma$ encodes the uncertainty in our approximate prediction for the bottleneck.

As a hint, note that the numerical solution of the BVP defined by Eqs.~\eqref{eq:geodesic} and \eqref{eq:approx_gamma} becomes much faster for low values of $\tilde\Gamma_0$. Accordingly, to obtain schedules as shown in Fig.~\ref{fig:schedules_N_17_seed_9129}, we usually start by finding a ``shallow'' schedule for $\tilde\Gamma_0=1.00$, which we then use as an initial guess for the next BVP at slightly increased $\tilde\Gamma_0$. This is repeated until we obtain smooth schedules at the final value of $\tilde\Gamma_0$.

\section{\label{sec:instances} Optimization Problems}

The strength of provable approximation bounds, e.g., as based on semi-definite programming~\cite{makarychevApproximationAlgorithmNonBoolean2012}, lies in their full generality: they apply to \textit{any} input, in particular to worst-case scenarios (the very hardest instances). In practice, heuristics can vastly outperform such approximation bounds on data sets of \textit{typical} instances, which are often mostly ``easy''. In looking for a quantum advantage in optimization, there are hence two distinct strategies: aiming for improvements on typical vs.\ on exceptionally hard instances. Since even the best classical algorithms can be expected to struggle on the latter (otherwise they might be capable of beating proven bounds), we consider making sure that quantum optimization algorithms perform as well as possible on hard instances to be the more promising strategy. 

To demonstrate the usefulness of our method as detailed in section~\ref{sec:geom_sched}, we work with two different sets of very hard instances. The first is comprised of the hardest random SK instances from the data of Ref.~\cite{bodeAdiabaticBottlenecksQuantum2024}; the details are given in section~\ref{subsec:sk_data} below. The second data set is taken from Shor \textit{et al.}~\cite{crossonDifferentStrategiesOptimization2014} and contains hard MAX 2-SAT instances at $N=20$. Note that this data set was also investigated in Ref.~\cite{mirkarimiComparingHardnessMAX2023}, where it was found that these instances are also harder than typical for a classical SAT solver. More the details are discussed below in section~\ref{subsec:max2sat_data}. 

Since all of the instances in the two data sets have been selected for in the sense of being hard for \textit{linear} annealing schedules at finite time, our main aim is to compare these to the one-component \textit{adaptive} schedules described in section~\ref{subsec:adap_sched}. We remark that selecting for hardness in this \textit{linear} sense is not unique, i.e.\ it is to be expected that, for example, some non-linear, two-component schedule might \textit{not} struggle to solve a given instance of this type. This is not a relevant concern, however, since it is just as possible to select for hardness in this ``two-component sense'', to which our general method then equally applies. 

\begin{figure}[htb!]
\begin{center}
\includegraphics[width=\linewidth]{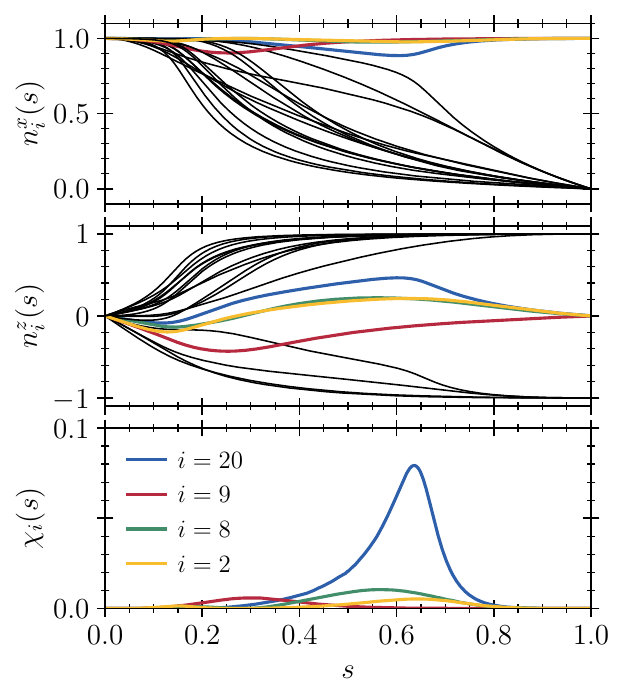}\vspace{-0.5cm}
\caption{\label{fig:semiclassical_crosson_N_20_idx_8} Example mean-field trajectories for one of the 130 hard MAX 2-SAT instances from Shor \textit{et al.}~\cite{crossonDifferentStrategiesOptimization2014}. The lowest panel shows the corresponding localization susceptibilities for the non-localizing spins, i.e.\ those spins with $n^z_i(1) \approx 0$. The adiabatic bottleneck is thus predicted to be located at $\tilde s_* = 0.62$. Running the corresponding adaptive annealing schedule at $T = 128$ ($T = 256$) results in an improvement factor of $P_{\mathrm{ad}} / P_{\mathrm{lin}} \approx  7.6\cdot 10^{3}\; (9.3\cdot 10^3)$.}
\end{center}
\end{figure}

We will denote the success probabilities in the linear and adaptive case, i.e.\ the probability of obtaining the optimal solution of a given instance with either type of schedule, throughout by $P_{\mathrm{lin}}$ and $P_{\mathrm{ad}}$, respectively. When taking ensemble averages of these quantities, we employ the \textit{geometric} mean
\begin{align}
    \bar P = \prod_{k=1}^{N_{\mathrm{ens}}}\sqrt[n]{P_k},
\end{align}
as this is the relevant average when asking for the success probability across an ensemble. Finally, we discuss how to relate our success probabilities $P$ to a cost measure like time-to-solution~\cite{albashAdiabaticQuantumComputation2018}. The total failure probability over $R$ runs of the quantum device will be $P_{\mathrm{fail}} = (1-P)^R$. Now, if we have some number $\mathcal{C}$ of approximate bottleneck \textit{candidates} (e.g.\ as in Fig.~\ref{fig:semiclassical_crosson_N_20_idx_8}) with corresponding adaptive schedules, where $P$ for one of them is much larger than for the rest, in the worst case we end up with at most the same $P_{\mathrm{fail}}$, yet over an increased number of runs $\mathcal{C}R$. Thus, one can define a lower bound for the effective success probability $\tilde P$ as
\begin{align}
    \tilde P \geq 1 - P_{\mathrm{fail}}^{1/\mathcal{C}R}.
\end{align}
We remark that, as in the case of Fig.~\ref{fig:semiclassical_crosson_N_20_idx_8} and~\ref{fig:mean_fields_N_17_seed_9129}, it will usually be straightforward to identify the relevant spin and consequently to avoid running several schedules.

\subsection{\label{subsec:sk_data} Sherrington-Kirkpatrick}

We take the hardest $N_{\mathrm{ens}}$ instances from the SK data set of Ref.~\cite{bodeAdiabaticBottlenecksQuantum2024}, which were originally selected by only retaining random instances with $\Delta < 0.01$. The ensemble sizes for the different system sizes $N$ are given in Tab.~\ref{tab:full_ens}. The instances in this data set have local fields $h_i$ and interactions $J_{ij}$ sampled from a standard normal distribution.

To extend our scaling results to larger system sizes, to these data we add 20 hard instances at $N=22$ that have been found by the following procedure: we run the mean-field evolution for a large number of random instances ($\sim 20000$) and retain only those for which the final effective magnetization of at least one spin is close to zero (the reasoning behind this is that we expect frustration and hardness to be correlated); then we run linear AQC for $T=2^7$ and keep those instances for which the success probability is below half a percentage (similar to the strategy Shor \textit{et al.}~\cite{crossonDifferentStrategiesOptimization2014}). As is indicated by our results in section~\ref{sec:results}, this procedure seems to produce even harder instances than those contained in { our previous SK data set~\cite{bodeAdiabaticBottlenecksQuantum2024}, where the selection of hard instances was based entirely on the size of the spectral gap obtained from exact diagonalization of the quantum Hamiltonian.} 

\begin{table}[htb!]
\centering
\def\arraystretch{1.5}
\begin{tabular}{c|c|c|c|c|c|c|c}
$N$ & 8 & 10 & 12 & 14 & 16 & 18 & 22 \\
\hline
$N_{\mathrm{ens}}$ & 250 & 250 & 250 & 250 & 250 & 110 & 20\\
\end{tabular}
\caption{Ensemble sizes for the hard SK data set. { As the computational cost for the largest system size $N=22$ is already considerable, we are limited to a relatively small number of instances.}}\label{tab:full_ens}
\end{table}

\subsection{\label{subsec:max2sat_data} Maximum 2-Satisfiability}

The MAX 2-SAT data from Shor \textit{et al.}~\cite{crossonDifferentStrategiesOptimization2014} were selected from among more than $2\cdot 10^5$ instances by retaining only those that had a linear-schedule success probability $P_{\mathrm{lin}} < 10^{-4}$ for a final time $T=100$. This originally led to 137 hard instances at $N=20$. However, we found six duplicates among these. Out of the remaining 131, we then also discard the instance ``easiest'' at our lowest $T=2^7$ to reach a final ensemble size of 130. Each instance has $N=20$ variables and 60 logical clauses, i.e.\ a clause-to-variable ratio of three, which is well beyond the hardness phase transition at the critical clause-to-variable ratio of one. Note that for these instances, we purposely did not calculate the exact spectrum, i.e.\ the exact bottlenecks $s_*$ are not known to us.

To be self-contained, we consider the following simple example of a MAX 2-SAT instance with four variables and six clauses:
\begin{align}
    \begin{split}
    C &= (x_1 \lor x_4) \land (\neg x_1 \lor x_4) \land (x_2 \lor \neg x_4) \\
      &\land (x_1 \lor \neg x_3) \land (x_1 \lor \neg x_2) \land (\neg x_2 \lor \neg x_3).        
    \end{split}
\end{align}
One can see right away from the first two clauses that $x_4$ needs to be true, and that, from the third clause, $x_2$ is consequently also true. This leads to $x_1$ being true and finally $x_3$ being false. Problems of this form can be mapped to an Ising Hamiltonian via
\begin{align}
    (x_i \lor \neg x_j) &\to \frac 14 \left(1 + \hat Z_i - \hat Z_j - \hat Z_i \hat Z_j\right),
\end{align}   
and analogously for the three remaining clause types. The total problem Hamiltonian is then simply the sum over all terms occurring in the conjunction.

\section{\label{sec:results} Results}

To provide experimental context for our simulation results, consider that current quantum annealers~\cite{kingQuantumCriticalDynamics2023} have dimensionless products of initial transverse-field strengths, $h_X$, and \textit{coherent} final times, $T$, of at most about 
\begin{align}
    h_X T \sim 1000.
\end{align}
Digitizing such analog quantum evolution by Trotterization into many QAOA layers via Eq.~\eqref{eq:qaoa_angles}, which then have to be run on a digital device, is likely still challenging even at present-day improved fidelities. For this reason, in units where $h_X=1$, the maximal final time we consider is $T=1024$. To obtain an order-of-magnitude intuition for the success probabilities we encounter below, consider a typical low linear-schedule success probability of $P_{\mathrm{lin}} = 10^{-6}$. Then even $R=10^4$ runs of the quantum device will result in a failure probability (i.e.\ the probability of \textit{not} seeing the optimal solution) of about $(1 - P_{\mathrm{lin}})^{R} \approx 99\%$. Since $P_{\mathrm{lin}}$ is bound to become worse for larger systems sizes $N$, one can already see from this simple back-of-the-envelope estimate why it is difficult to optimize annealing schedules { for maximal success probability} just by trial and error: if the ground state is ``invisible'' for some finite $R$, then it will effectively not influence the optimization. {
Note, however, that a choice of cost function other than the success probability might not suffer from this limitation.}

\begin{figure*}[htb!]
\begin{center}
\includegraphics[width=\linewidth]{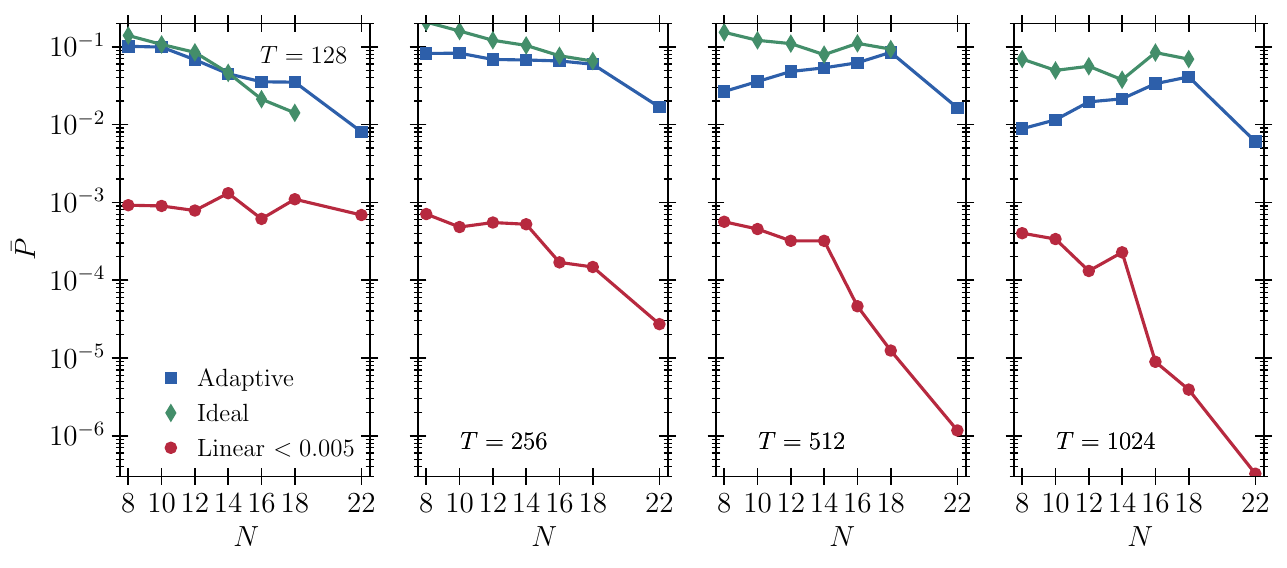}\vspace{-0.5cm}
\caption{\label{fig:all_ps} Geometric mean success probability $\bar P$ for different systems sizes $N$ and final times $T$. For each $N$ and $T$, we retain only those instances that have $P < 0.5\%$ in the linear case; the corresponding effective ensemble sizes are given in Tab.~\ref{tab:eff_ens}. The ``adaptive'' schedules are derived from the semi-classical prediction $\tilde s_*$, while the ``ideal'' schedules are based on the exact bottleneck position $s_*$ defined in Eq.~\eqref{eq:bottleneck}. The parameters of Eq.~\eqref{eq:approx_gamma} used for both the ideal and adaptive schedules are $\tilde\Gamma_0 = 3.20$ and $\sigma = 0.05$. The Trotter step size is $T/p=2^{-3}$ throughout. Note that the instances at $N=22$ have lower $P$ because these were generated by a different procedure than was used for the smaller systems sizes (cf.~section~\ref{sec:instances}).}
\end{center}
\end{figure*}

\subsection{Sherrington-Kirkpatrick}\label{subsec:SK_results}

\begin{figure}[htb!]
\begin{center}
\includegraphics[width=\linewidth]{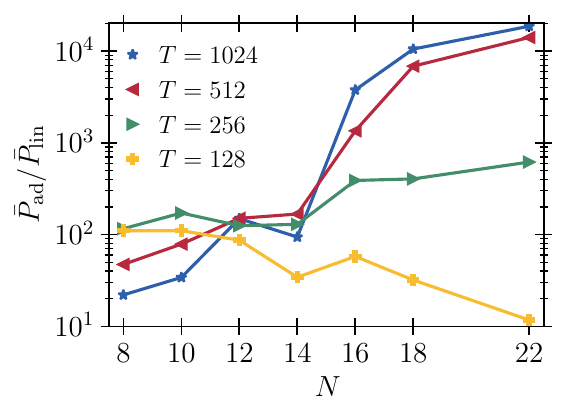}\vspace{-0.5cm} 
\caption{\label{fig:p_adap_p_lin} Geometric means across the SK ensembles of the improvements $P_{\mathrm{ad}}/P_{\mathrm{lin}}$ for different final times $T$.}
\end{center}
\end{figure}

Our main results for the SK data set { are} shown in Fig.~\ref{fig:all_ps} and~\ref{fig:p_adap_p_lin}. The effective ensemble sizes after imposing a success-probability cutoff $P<0.5\%$ are given in Tab.~\ref{tab:eff_ens}. We use this cutoff to make sure our analysis is focused on instances that are actually hard for the linear schedules at each $T$. Since it is numerically very expensive to find such instances, the ensemble sizes are moderate.

\begin{table}[htb!]
\centering
\def\arraystretch{1.5}
\begin{tabular}{c|ccccccc}
$N$ & 8 & 10 & 12 & 14 & 16 & 18 & 22 \\
\hline
$T=2^7$ & 33 & 41 & 44 & 25 & 46 & 19 & 19 \\
\hline
$T=2^8$ & 96 & 88 & 94 & 69 & 88 & 41 & 20 \\
\hline
$T=2^9$ & 144 & 125 & 131 & 102 & 128 & 57 & 20 \\
\hline
$T=2^{10}$ & 156 & 144 & 129 & 119 & 152 & 68 & 19
\end{tabular}
\caption{Effective ensemble sizes for the hard SK data set when retaining only those instances with a linear success probability $P_{\mathrm{lin}} < 0.5\%$ at different final times $T$.}\label{tab:eff_ens}
\end{table}

The ``ideal'' results shown in both Fig.~\ref{fig:all_ps} and~\ref{fig:p_adap_p_lin} are calculated by inserting the \textit{exact} bottleneck position $s_*$ into Eq.~\eqref{eq:approx_gamma}, whereas the approximate $\tilde s_*$ is used for the ``adaptive'' results; for example schedules, s.~Fig.~\ref{fig:schedules_N_17_seed_9129}. Note also that we only know the exact bottleneck $s_*$ to a relatively low precision of $2^{-5}$, i.e.\ one should not expect our ``ideal'' results to be fully ideal; for $N=22$, we did not attempt to compute the exact $s_*$ and hence do not have the corresponding ``ideal'' data point.

One of the most striking features of Fig.~\ref{fig:all_ps} is that the geometric mean $\bar P$ across the ensembles goes down systematically with $T$ when using a linear schedule. { As discussed in detail in Appendix~\ref{app:diab_enhance}}, we do not conclude from this that running for shorter times is a valid strategy. A related obvious question to ask is whether there is anything to learn from the fact that the decrease in $\bar P$ is \textit{stronger} for larger $N$? One hypothesis consistent with the data appears to be this: the smaller instances (i.e.\ $N=8, ..., 14$) are, in fact, evolving almost adiabatically already for $T=128$, while the larger (and statistically harder) instances cross over into the adiabatic regime from left to right. Given Fig.~\ref{fig:SK_linear_ex_states}, we argue that the results we see for $N > 14$ in the two right panels should inform our generic expectations: in an energy landscape with many local minima far away (in Hamming distance) from the optimum, a return to the ground state after excitation cannot be relied upon.

Another feature of Fig.~\ref{fig:all_ps} to comment on is that the ``adaptive'' results outperform the ``ideal'' for $N=16, 18$ at $T=128$. Again, we attribute this to spurious diabatic effects. For intermediate $T$ (second and third panel), the fact that the two curves are very close may be taken as further evidence that the localization susceptibility $\chi_i(s)$ does indeed predict the bottlenecks well.  Finally, we remark that the new hard instances at $N=22$ were obtained with a different procedure (cf.~\ref{subsec:sk_data}) that seems to result in even harder instances than the generation method used in Ref.~\cite{bodeAdiabaticBottlenecksQuantum2024}. While the corresponding kink in the ``adaptive'' curves seems to indicate that our method becomes worse for these instances, we point to Fig.~\ref{fig:p_adap_p_lin}, which shows the success improvement relative to the linear case. The improvements of more than four orders of magnitude are rather striking.

\subsection{Maximum 2-Satisfiability}\label{subsec:MAX2SAT_results} 

\begin{figure}[htb!]
\begin{center}
\includegraphics[width=\linewidth]{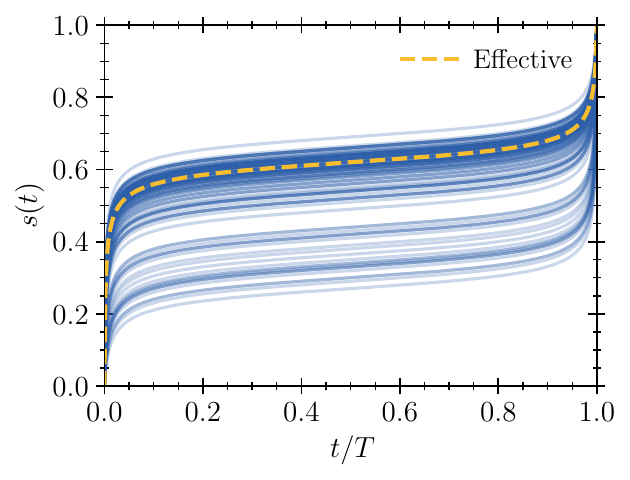}\vspace{-0.5cm}
\caption{\label{fig:crosson_schedules} The collection of all adaptive schedules we found for the 130 hard MAX 2-SAT instances from Shor \textit{et al.}~\cite{crossonDifferentStrategiesOptimization2014}. In most cases, the adiabatic bottleneck is located around $\tilde s_* \approx 0.60$. The parameters of Eq.~\eqref{eq:approx_gamma} used in all cases are $\tilde\Gamma_0 = 3.20$ and $\sigma = 0.05$. We remark that it is straightforward to change these parameters to obtain either shallower or steeper schedules. }
\end{center}
\end{figure}

While the results on the SK instances are very promising, one might argue that the SK model is bound to benefit from mean-field models by its very nature. To guard against this possibility, we now move on to the investigation of the hard MAX 2-SAT instances from Shor \textit{et al.}~\cite{crossonDifferentStrategiesOptimization2014}, which have very uniform local fields $h_i$ and interactions $J_{ij}$. An exemplary set of mean-field trajectories was already shown above in Fig.~\ref{fig:semiclassical_crosson_N_20_idx_8}. Note that it is typical for MAX 2-SAT to possess a greater number of non-localizing spins, i.e.\ spins that have $n_i^z(1)\approx 0.0$. Interestingly, while this means that the mean-field approximate optimization algorithm~\cite{misra-spieldennerMeanFieldApproximateOptimization2023} results in a {coin toss} for these spins in particular, their semi-classical fluctuations contain useful information about the adiabatic bottleneck (cf.~lowest panel of Fig.~\ref{fig:semiclassical_crosson_N_20_idx_8}).

Proceeding as before, for each instance we now use the localization susceptibilities $\chi_i(s)$ of the frustrated spins to obtain a semi-classical prediction $\tilde s_*$ for the adiabatic bottleneck. The resulting ``adaptive'' schedules are presented in Fig.~\ref{fig:crosson_schedules} for the entire ensemble of 130 instances. The schedules have ``steep'' sections at the beginning and end, yet are slowly and smoothly changing at intermediate times, which is in excellent agreement with the general constraints on schedules derived from optimal-control theory in Ref.~\cite{bradyOptimalProtocolsQuantum2021} 

Note that we also derive an ``effective'' schedule with $\tilde s_* = 0.62$ from the observation that the slowing down happens around this point for the majority of instances (dashed yellow line in Fig.~\ref{fig:crosson_schedules}). We take this as an example of how our semi-classical tools can be used to study an ensemble of instances as a whole to learn useful collective properties. 

\begin{figure}[htb!]
\begin{center}
\includegraphics[width=\linewidth]{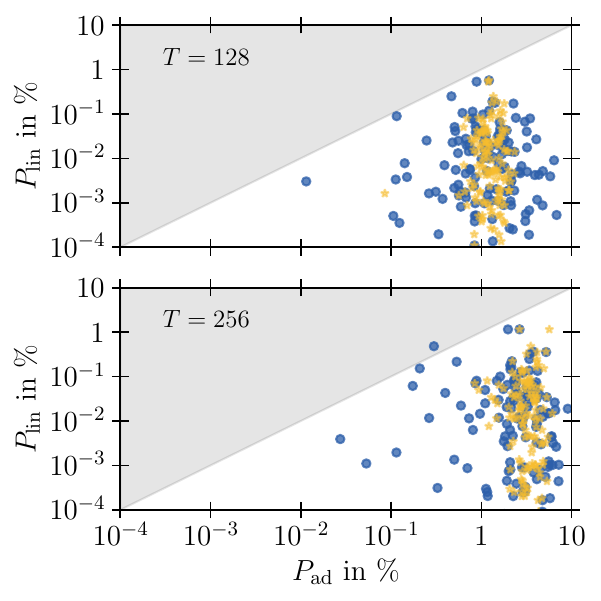}\vspace{-0.5cm}
\caption{\label{fig:crosson_improvement} Success probability $P_{\mathrm{lin}}$ vs.\ $P_{\mathrm{ad}}$ (blue circles) for two different final times. The yellow stars plotted across are the respective results when using the ``effective'' schedule from Fig.~\ref{fig:crosson_schedules}. We use Trotter steps of size $T/p=2^{-4}$ in the upper and  $T/p=2^{-3}$ in the lower panel.}
\end{center}
\end{figure}

\begin{figure}[htb!]
\begin{center}
\includegraphics[width=\linewidth]{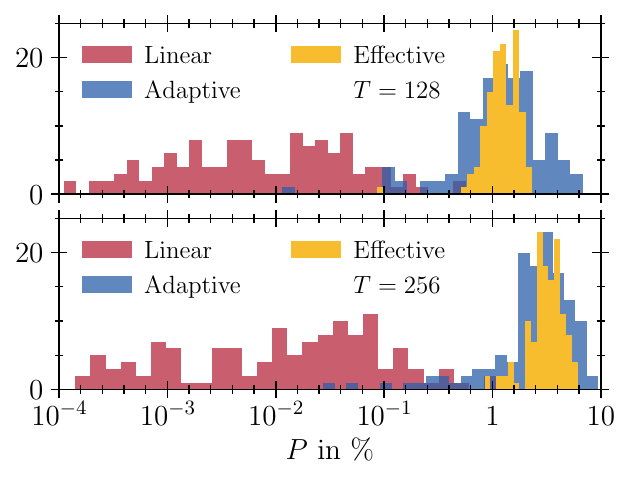}\vspace{-0.5cm}
\caption{\label{fig:crosson_hist} Histograms detailing the same results as shown in Fig.~\ref{fig:crosson_improvement}. Observe the { reduced} variance in the ``effective'' results, an explanation for which is provided in the main text. For $T=128$ (upper panel), the geometric mean success probabilities are $\bar P_{\mathrm{lin}} = 0.007\%$, $\bar P_{\mathrm{ad}} = 1.1\%$ and $\bar P_{\mathrm{eff}} = 1.2\%$ for the linear, adaptive and effective schedules, respectively; for $T=256$ (lower panel), we instead find $\bar P_{\mathrm{lin}} = 0.01\%$, $\bar P_{\mathrm{ad}} = 2.3\%$ and $\bar P_{\mathrm{eff}} = 3.1\%$. }
\end{center}
\end{figure}

The main results of this section, obtained from running AQC (or, for that matter, fixed-angle QAOA) simulations for all instances with the three different types of schedules, are given in Fig.~\ref{fig:crosson_improvement} and~\ref{fig:crosson_hist}. Note again that we employ a second-order Trotter-Suzuki expansion~\cite{willschBenchmarkingQuantumApproximate2020} for the linear schedules. We have also verified for $T=128$ that lowering the step size from $T/p = 2^{-3}$ to $2^{-4}$ does indeed slightly \textit{improve} our adaptive results, i.e.\ the results obtained at the former step size for $T=256$ should be conservative.

The first thing to notice about Fig.~\ref{fig:crosson_improvement} is that our linear success probabilities $P_{\mathrm{lin}}$ at both values of $T$ are higher than the values quoted by Shor \textit{et al.}~\cite{crossonDifferentStrategiesOptimization2014} at $T=100$; for this final time they found $10^{-5}\% \leq P_{\mathrm{lin}} \leq 10^{-2}\%$ across the ensemble. Therefore, while our $P_{\mathrm{lin}}$ remain low, there is an increase after all for evolution times that are more adiabatic; also, since our results are not expected to become worse for shorter times, this observation means that our improvement factors are probably rather conservative. Even so, in Fig.~\ref{fig:crosson_improvement} we only find a small number of instances close to the diagonal, while the bulk of them sit at considerable improvements. 

Perhaps most interesting is the good performance of the ``effective'' schedule from Fig.~\ref{fig:crosson_schedules}, which seems to confirm that it is possible to roughly estimate the most typical adiabatic bottleneck across the entire ensemble from semi-classical information.

While the linear results hardly improve from top to bottom in Fig.~\ref{fig:crosson_hist}, the adaptive results visibly shift towards higher success probabilities. Since Shor \textit{et al.}~\cite{crossonDifferentStrategiesOptimization2014} also did \textit{not} observe an increase in the linear success probability for \textit{larger} final times than $T=100$, it seems possible to interpret the fact that our results \textit{do} improve from top to bottom as an indicator that we indeed predict the true bottleneck reasonably accurately. The lower spread in the ``effective'' results, stemming from the corresponding schedule shown in Fig.~\ref{fig:crosson_schedules}, is also worth commenting on: apparently, the instances ending up in the left tails of the ``adaptive'' distributions have indeed received below-average $\tilde s_*$ predictions that are then { improved} by using { the aggregate value} $\tilde s_* = 0.62$ instead; however, it would appear that the opposite happens on the right tail where presumably very accurate predictions are shifted towards lower probabilities.

In the original work on the MAX 2-SAT data we use here, Shor \textit{et al.}~\cite{crossonDifferentStrategiesOptimization2014} obtained good improvements of the success probability by using catalyst Hamiltonians made from random Pauli words of length one and two. It should be relatively straightforward to include such terms into the mean-field Hamiltonian~\eqref{eq:mf_H} and the paramagnon Hamiltonian~\eqref{eq:para_ham} to the effect of \textit{combining} both methods to obtain even greater improvements.

\section{Conclusion}\label{sec:conc}

The results we have presented in this work integrate tools from different disciplines to yield, at best, orders-of-magnitude improvements in the success probabilities of hard instances. The study of dynamic correlation functions such as Eq.~\eqref{eq:stat_func}, in particular, appears very natural as an approach to first-order quantum phase transitions. Note that disorder correlations have been investigated recently by Nishimori~\cite{nishimoriAnomalousDistributionMagnetization2024} in the context of Ising spin glasses; we point out that it is straightforward, within our spin coherent-state path integral, to move beyond the ``diagonal'' paramagnon fluctuations in Eq.~\eqref{eq:paramagnon_number} and investigate dynamic spin-spin correlation functions. Since these semi-classical methods are not limited to small systems, one can consider them as next-to-leading-order tools to shine more light onto the formidable black box represented by large-scale spin glasses. 

{ Note also that, while solving the embedding problem, i.e.\ the problem of mapping a given discrete optimization problem onto an existing quantum hardware graph, is a challenge in itself~\cite{pelofskeComparingThreeGenerations2023, fangMinimizingMinorEmbedding2020}, our tools are embedding-agnostic and can be leveraged to improve the performance arbitrary hardware configurations.}

{ Further interesting avenues could be found in applying our methods to improve alternative combinatorial-optimization frameworks such as simulated bifurcation machines~\cite{gotoCombinatorialOptimizationSimulating2019, gotoHighperformanceCombinatorialOptimization2021, tatsumuraEnhancingInvehicleMultiple2024} and digital annealing~\cite{munchTransformationDependentPerformanceEnhancementDigital2023, schonbergerQuantumInspiredDigitalAnnealing2023}.}

We close by remarking once more that our methods can be \textit{combined} with the application of catalysts~\cite{crossonDifferentStrategiesOptimization2014, ghoshExponentialSpeedquantumAnnealing2024}, which are a promising tool to mitigate exponentially small spectral gaps; achieving a ``double'' improvement in this way appears to be an interesting direction for future work.

\begin{acknowledgments}
The authors acknowledge partial support from the German Federal Ministry of Education and Research, under the funding program ``Quantum technologies - from basic research to the market'', Contract Numbers 13N15688 (DAQC), 13N15584 (Q(AI)2) and from the German Federal Ministry of Economics and Climate Protection under contract number, 01MQ22001B (Quasim). 
\end{acknowledgments}

\appendix

\subsection{Is diabatic enhancement a finite-size effect?}\label{subsec:diabatic_finite_size}\label{app:diab_enhance}

To provide additional context for our main results in sections~\ref{subsec:SK_results} and~\ref{subsec:MAX2SAT_results}, here we briefly discuss an observation that was already made by Shor \textit{et al.}~\cite{crossonDifferentStrategiesOptimization2014}, namely that \textit{shorter} evolution times $T$, in fact, seem to \textit{increase} the success probability $P$, which they referred to as ``The hare beats the tortoise''. The obvious question to ask is whether this is simply a finite-size effect at small $N$.

\begin{figure}[htb!]
\begin{center}
\includegraphics[width=\linewidth]{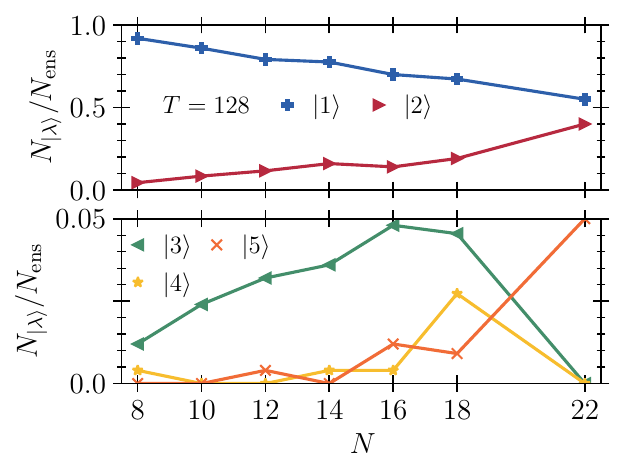}\vspace{-0.5cm}
\caption{\label{fig:SK_linear_ex_states} The number of excited states per SK ensemble to which linear AQC converges at $T=128$, i.e.\ the state $|\lambda\rangle$ with the highest success probability. Clearly, the ground state $|0\rangle$ quickly becomes more ``invisible'' as $N$ increases.}
\end{center}
\end{figure}

Shor \textit{et al.}~\cite{crossonDifferentStrategiesOptimization2014} tried to answer this question by applying diabatic evolution with a linear schedule to the AQC version of Grover's algorithm. Since the spectral gap in this case closes exponentially with the qubit number, they argued that diabatic enhancement should not work, which indeed they observed for $N=20$. However, their conclusion that this hints at diabatic enhancement \textit{not} being a finite-size effect seems to be a \textit{non sequitur}: we would expect hard instances of any problem at large $N$ to be just of this sort, i.e.\ to possess exponentially small spectral gaps. 

Feinstein \textit{et al.}~\cite{feinsteinRobustnessDiabaticEnhancement2024} give a detailed discussion of diabatic strategies to enhance the overlap of the annealed final state with the ground state. Their main conclusion is the existence of a trade-off in the required precision between the diabatic annealing time and the parameters controlling, e.g., a catalyst Hamiltonian (typically taken to be proportional to $s(1-s)$). That is, precise knowledge of the diabatic runtime required to maximize the final ground-state overlap (via excitation followed by decay) lowers the precision required in the design of the catalyst, and vice versa. In the absence of precise enough knowledge about either, the conclusion to be drawn is that diabatic evolution will, for hard instances, generically lead to an exponentially small overlap with the ground state, the only remedy for which is then to run the quantum optimization exponentially many times. As mentioned before, this point is one of the main motivations for our method to systematically increase the success probability via tailored schedules.

Another interesting perspective, however, is afforded by the fact that it is possible to apply our semi-classical theory to a Hamiltonian that \textit{includes} the ansatz for a catalyst. It is not inconceivable that this might help both in designing catalysts as well as yielding further improvements by combination with appropriate adaptive schedules.

To conclude this Appendix, we turn to Fig.~\ref{fig:SK_linear_ex_states}, which shows the ensemble statistics of the most likely final states $|\lambda\rangle$, computed for $T=128$ with the full SK ensembles as quoted in Tab.~\ref{tab:full_ens}. For one thing, the fraction of instances for which the ground state is most likely turns out to be very small (which is, of course, to be expected by construction). More strikingly, there is a clear trend with $N$ towards higher excited states. We believe this should be interpreted as a clear indicator that diabatic enhancement { becomes less effective for} large systems.

\bibliography{refs}

\end{document}